\documentclass[12pt]{article}
\usepackage{amsfonts}
\usepackage{amssymb}
\usepackage{amscd}

\begin{document}

\title{\large{\bf ADELIC  HARMONIC  OSCILLATOR}}

\bigskip

\author{ {\bf Branko Dragovich} \\ Institute of Physics, P. O. Box 57 \\ 11001 Belgrade, Yugoslavia}

\date{}
\bigskip

\bigskip

\bigskip

\maketitle

\begin{abstract}
{ Using the Weyl quantization we formulate one-dimensional adelic
quantum mechanics, which unifies and treats ordinary and $p$-adic
quantum mechanics on an equal footing. As an illustration the
corresponding harmonic oscillator is considered. It is a simple,
exact and instructive adelic model. Eigenstates are
Schwartz-Bruhat functions. The Mellin transform of a simplest
vacuum state leads to the well known functional relation for the
Riemann zeta function. Some expectation values are calculated. The
existence of adelic matter at very high energies is suggested.}
\end{abstract}

\bigskip

\bigskip
\noindent{\bf 1.\ \ Introduction}
\bigskip

Since 1987, the application of $p$-adic numbers has been of
interest in string theory \cite{volovich1}-\cite{vladimirov1},
quantum mechanics \cite{volovich2}- \cite{zelenov2}, and some
other areas of theoretical \cite{dragovich2} and mathematical
\cite{vladimirov2}- \cite{dragovich3} physics (for a review see
Refs. \cite{volovich6} and \cite{freund2}). Many $p$-adic models
have been constructed. In string theory a product (adelic) formula
is obtained, i.e. the product of the ordinary four-point amplitude
and all $p$-adic analogs is equal to a constant.

One of the significant achievements in this field is the
formulation of $p$-adic quantum mechanics
\cite{volovich2},\cite{volovich3},\cite{volovich4} (see also Refs.
\cite{alacoque} and \cite{meurice}). The corresponding model of
the harmonic oscillator is constructed, and its evolution and
spectral properties are analyzed \cite{zelenov1}-\cite{zelenov2}.
It is an exactly soluble model.

The connection of $p$-adic quantum mechanics with ordinary
mechanics has so far been an open problem. Problems of this kind
are more or less a characteristic of all other $p$-adic models.

In this article we use the concept of adeles as a basis to unify
$p$-adic and ordinary quantum mechanics. In particular, we
investigate the adelic model of the harmonic oscillator. Some
aspects of the adelic approach have been earlier considered in
$p$-adic string theory \cite{freund1}-\cite{vladimirov1} and
$p$-adic quantum field theory \cite{roth}. As we shall see, the
adelic harmonic oscillator is an exact model which exhibits many
interesting mathematical and physical properties.

In Sec. 2 we present some mathematics of $p$-adic numbers and
adeles (one can also see Refs. \cite{vladimirov2} and
\cite{gelfand}). Section 3 contains the ordinary and $p$-adic
harmonic oscillator in classical and quantum mechanics. In
formulation of $p$-adic quantum mechanics we follow the
Vladimirov-Volovich approach
\cite{volovich2},\cite{volovich3},\cite{volovich4}. This approach
is generalized to adelic quantum mechanics in Sec. 4. As an
example  the adelic model of the harmonic oscillator is
considered. The main mathematical and physical aspects of the
adelic quantum approach are discussed in the last section.

\bigskip
\noindent{\bf 2.\ \ On Adeles}
\bigskip

There is sense in starting with the field of rational numbers
${\mathbb Q}$. From the physical point of view, ${\mathbb Q}$
contains all experimental data. In mathematics, ${\mathbb Q}$ is
the simplest infinite number field. Mathematical models of
physical phenomena prefer completions (and algebraic extensions)
of ${\mathbb Q}$ rather than ${\mathbb Q}$ itself. Completions of
${\mathbb Q}$ with respect to to the usual absolute value $| \,\,
|_\infty$ and $p$-adic norm (valuation) $| \,\, |_p$ give the
field of real numbers ${\mathbb R} \equiv {\mathbb Q}_\infty$ and
the field of of $p$-adic numbers ${\mathbb Q}_p$  ($p$ = a prime
number), respectively. According to the Ostrowski theorem,
${\mathbb R}$ and ${\mathbb Q}_p \, \, \, (p = 2, 3, 5, \cdots)$
exhaust the possible number fields which can be obtained by
completions of ${\mathbb Q}$.

A series
$$
\varepsilon \sum_{k\in{\mathbb Q}} a_k p^k \, , \quad a_k \in \{
0, 1, \cdots, p-1\} \,, \eqno(2.1)
$$
where $\varepsilon =\pm 1$, and $a_k =0$ for $k \geq k_0$ is a
real number. If $\varepsilon = 1$ and $a_k = 0$ for $k \leq k_0$
the series (2.1) represents a $p$-adic number in ${\mathbb Q}_p$.
The ring of $p$-adic integers is ${\mathbb Z}_p = \{x \in {\mathbb
Q}_p \, : |x|_p \leq 1  \}$, i.e. for ${\mathbb Z}_p \,, \, \,
k\geq 0$ in (2.1).

Since $ {\mathbb R}$ and ${\mathbb Q}_p$ are locally compact
groups one can define on them two invariant measures. On the
additive groups  ${\mathbb R}^+$ and ${\mathbb Q}_p^+$ there are
translationally  invariant  measures $d x_\infty$ and $d x_p$,
respectively, where $d x_\infty$ corresponds to the Lebesque
measure normalized by $\int_{|x_\infty|_\infty \leq 1} d x_\infty
=2$, and  $d x_p$ is the Haar measure normalized by $\int_{|x_p|_p
\leq 1} d x_p =1$. On the multiplicative groups ${\mathbb R}^\ast$
and ${\mathbb Q}_p^\ast$ there exist the Haar measures $d^\ast
x_\infty$ and $d^\ast x_p$, respectively, invariant under
multiplication. These two types of measure are connected by the
equalities
$$
d^\ast x_\infty =  \frac{d x_\infty}{|x_\infty|_\infty} \,, \qquad
d^\ast x_p = \frac{1}{1 - p^{-1}}\, \frac{d x_p}{|x_p|_p} \, .
\eqno(2.2)
$$

It is significant that real and $p$-adic numbers can be unified by
means of adeles. An adele is an infinite sequence
$$
   a = (a_\infty , a_2 , \cdots , a_p \,, \cdots) \, , \eqno(2.3)
$$
where $a_\infty \in {\mathbb R}\,$ and $p$-adic numbers $ a_p \in
{\mathbb Z}_p $ for all but a finite number of $p$. The set of
adeles ${\mathcal A}$   may be regarded as a direct topological
product $ {\mathbb Q}_\infty  \times \prod_p {\mathbb Q}_p $ whose
elements satisfy the above restriction. ${\mathcal A}$ is a ring
under componentwise addition and  multiplication. Denote by
${\mathcal A}^+$ the additive group ${\mathcal A}$.
 A multiplicative group of ideles ${\mathcal A}^\ast$ is a subset
of ${\mathcal A}$ with elements $b = (b_\infty , b_2 , \cdots ,
b_p \,, \cdots)$ such that $b_\infty \neq 0$  and $b_p \neq 0$ for
every $p$, and $|b_p|_p = 1$ for all but a finite number of $p$. A
principal adele (idele) is a sequence $(r, r, \cdots , r , \cdots)
\in {\mathcal A}$ , where $r \in {\mathbb Q}\, \, \, \,\, (r \in
{\mathbb Q}^\ast  = {\mathbb Q} \setminus \{ 0 \})$. One can
define a product of norms for  adeles,
$$
|b| = |b_\infty|_\infty \,  \prod_p \,  |b_p|_p \,\,,  \eqno(2.4)
$$
which for a principal idele is
$$
|r| = |r|_\infty \,  \prod_p \,  |r|_p  = 1 .  \eqno(2.5)
$$

An additive character on ${\mathcal A}^+$ is
$$
\chi (x y) = \chi_\infty (x_\infty y_\infty)\, \prod_p \, \chi_p
(x_p y_p) = \exp (- 2 \pi i x_\infty y_\infty) \, \prod_p \exp{
 2 \pi i \{  x_p y_p\}}_p  \, ,       \eqno(2.6)
$$
where $x,\, y \in {\mathcal A}^+$ and $\{ a_p\}_p$ is the
fractional part of $a_p \in {\mathbb Q}_p$. A multiplicative
character on ${\mathcal A}^\ast$ can be defined as
$$
\pi (b) =  \pi_\infty (b_\infty)\, \pi_2 (b_2) \cdots \pi_p (b_p)
\cdots = |b_\infty|_\infty^s \, \prod_p |b_p|_p^s = |b|^s \, ,
\eqno(2.7)
$$
where $b$ is an idele and $s $ is a complex number. In fact, only
finitely many factors in (2.6) and (2.7) are different from unity.
One can show that $\chi (r) =1 $ if $r$ is a principal adele, and
$\pi (r) = 1$ if $r$ is a principal idele.

An elementary function on the group of adeles ${\mathcal A}^+$ is
$$
\varphi (x) =\varphi_\infty (x_\infty)\, \prod_p   \varphi_p
(x_p)\, ,                  \eqno(2.8)
$$
where $x\in {\mathcal A}^+ \, , \, \, \, \varphi_\infty (x_\infty)
\in {\mathcal S}({\mathbb R})\, , \, \varphi_p (x_p) \in {\mathcal
S}({\mathbb Q}_p) $. Namely, $ \varphi (x)$ is a complex-valued
function which satisfies the following conditions: ({\it i})
$\varphi_\infty (x_\infty)$ is an infinitely differentiable
function on ${\mathbb R}$ and   $|x_\infty|_\infty^n \,\,
\varphi_\infty (x_\infty) \rightarrow 0$ as  $ |x_\infty|_\infty
\rightarrow \infty $  for any $n\in \{ 0, 1, 2 , \cdots \}$; ({\it
ii}) $\varphi_p (x_p)$ is a finite and locally constant function,
i.e. $\varphi_p$ has a compact support and $\varphi_p (x_p + y_p)
= \varphi_p (x_p)$ if $|y_p|_p\leq p^{-n} \, ,\, n=n (\varphi_p) $
; and ({\it iii}) $ \varphi_p (x_p) = \Omega (|x_p|_p) ,$ for all
but a finite number of $p$ ,where
$$
\Omega (u) = \left\{  \begin{array}{ll}
                 1,   &  \mbox{if} \,\,\,\, 0\leq u \leq 1,  \\
                 0,   &  \mbox{if} \,\,\,\,  u > 1 .
                 \end{array}    \right.   \eqno(2.9)
$$
All finite linear combinations  of elementary functions (2.8) make
up the  set ${\mathcal S}({\mathcal A})$ of the Schwartz-Bruhat
functions.

The Fourier transform of $\varphi (x) \in {\mathcal S}({\mathcal
A})$ is
$$
\tilde{\varphi} (y) = \int_{{\mathcal A}^+} \varphi (x)\, \chi (x
y) \, dx = \int_{{\mathbb R}} \varphi_\infty (x_\infty) \, \,
\chi_\infty( x_\infty \, y_\infty) \, dx_\infty
$$
$$
\times \prod_p \int_{{\mathbb Q}_p} \varphi_p (x_p) \,\, \chi_p
(x_p\,  y_p )\, \, dx_p \, ,     \eqno(2.10)
$$
where $ dx = dx_\infty \, dx_2 \cdots dx_p \cdots$ is the Haar
measure on  ${\mathcal A }^+$. The Fourier transform maps
${\mathcal S}({\mathcal A})$ onto ${\mathcal S}({\mathcal A})$.
The Mellin transform of $ \varphi (x) \in {\mathcal S}({\mathcal
A})$ is defined with respect to the multiplicative character $\pi
(x) = |x|^s$ :
$$
\Phi (s) = \int_{{\mathcal A}^\ast} \varphi (x)\, |x|^s \, d^\ast
x =  \int_{{\mathbb R}} \varphi_\infty (x_\infty) \, \,
|x_\infty|_\infty^{s-1} \, dx_\infty
$$
$$
\times \prod_p \int_{{\mathbb Q}_p} \varphi_p (x_p) \,\,
|x_p|_p^{s-1}\, \, \frac{ dx_p }{1 - p^{-1}}\, , \,\, \, \,
\mbox{Re}\,s
> 1 \, ,\eqno(2.11)
$$
where $ d^\ast x = d^\ast x_\infty \, d^\ast x_2 \cdots d^\ast x_p
\cdots$ is the Haar measure on  ${\mathcal A }^\ast$.

The function $\Phi (s)$ can be  analytically continued on the
whole field of complex numbers, except $s=0$ and $s=1$, where it
has simple poles with residue $- \varphi (0)$ and $\tilde{\varphi}
(0)$, respectively. Let $\Phi (s)$ is the Mellin transform of
$\tilde{\varphi}$. One can show \cite{gelfand} that $\Phi$ and
$\tilde{\Phi}$ are connected by the Tate formula
$$
\Phi (s) = \tilde{\Phi} (1-s)\, .    \eqno(2.12)
$$

Recall that the Hilbert space $H$ is a set such that: $(\it i)$
$H$ is an infinitely dimensional linear system; $(\it ii)$ there
is a scalar product $(f,\, g)$ for any $f,\, g \in H$; and $(\it
iii)$ H is a complete space with respect to the metric $d (f,\, g)
=\parallel f - g\parallel$, where the norm $\parallel f\parallel =
(f,\, f)^{\frac{1}{2}}$. We use here the following Hilbert spaces.

$(1)\, \,$  $L_2 ({\mathbb R})$ is a space of complex-valued
functions $\psi_1^{(\infty)} (x_\infty) , \psi_2^{(\infty)}
(x_\infty) , \cdots $ with scalar product and norm
$$
(\psi_1^{(\infty)}  ,\, \psi_2^{(\infty)} )  = \int_{{\mathbb R}}
\bar{\psi}_1^{(\infty)} (x_\infty) \, \psi_2^{(\infty)} (x_\infty)
\, dx_\infty \, , \eqno(2.13a)
$$
$$
\parallel \psi^{(\infty)} \parallel = (\psi^{(\infty)}, \,
\psi^{(\infty)})^{\frac{1}{2}}  < \infty \, ,  \eqno(2.13b)
$$
where $x_\infty \in {\mathbb R}$ and $dx_\infty$ is the Lebesgue
measure.

$(2)\, \,$  $L_2 ({\mathbb Q}_p)$ is a space of complex-valued
functions $\psi_1^{(p)} (x_p) , \psi_2^{(p)} (x_p) , \cdots $ with
scalar product and norm
$$
(\psi_1^{(p)}  ,\, \psi_2^{(p)} )  = \int_{{\mathbb Q}_p}
\bar{\psi}_1^{(p)} (x_p) \, \psi_2^{(p)} (x_p) \, dx_p \, ,
\eqno(2.14a)
$$
$$
\parallel \psi^{(p)} \parallel = (\psi^{(p)}, \,
\psi^{(p)})^{\frac{1}{2}}  < \infty \, ,  \eqno(2.14b)
$$
where $x_p \in {\mathbb Q}_p$ and $dx_p$ is the Haar measure on
${\mathbb Q}_p$.

$(3)\, \,$  $L_2 ({\mathcal A})$ is a space of complex-valued
functions $\psi_1 (x) , \psi_2 (x) , \cdots $ with scalar product
and norm
$$
(\psi_1  ,\, \psi_2 )  = \int_{{\mathcal A}} \bar{\psi}_1 (x) \,
\psi_2 (x) \, dx \, , \eqno(2.15a)
$$
$$
\parallel \psi \parallel = (\psi , \,
\psi )^{\frac{1}{2}}  < \infty \, ,  \eqno(2.15b)
$$
where $x \in {\mathcal A}$ and $dx $ is the Haar measure on
${\mathcal A}$.

$L_2 ({\mathbb R})$,  $L_2 ({\mathbb Q}_p)$ and  $L_2 ({\mathcal
A})$ are the separable Hilbert spaces.

The bases of the above spaces may be given by the orthonormal
eigenfunctions of some operator (e.g. an evolution operator). Let
the bases for the evolution operator in the real $[U_\infty
(t_\infty)]$ and $p$-adic $[U_p (t_p)]$ cases be $\{\psi_{n
m}^{(\infty)}\} $ and $\{ \psi_{\alpha_p \beta_p}^{(p)}  \}$,
where $n, m = 0, 1, \cdots $, and $\alpha_p \, , \beta_p$ are
indices which characterize energy and its degeneration. Also, let
us denote by $\psi_{0 \beta_p}^{(p)}$ (vacuum states)
eigenfunctions which are invariant under $U_p (t_p)$. We define an
orthonormal basis for the corresponding adelic evolution operator,
$$
U (t) = U_\infty (t_\infty) \prod_p U_p (t_p) \, , \quad t \in
{\mathcal A} \, , \eqno(2.16)
$$
as
$$
\psi_{\alpha \beta} (x) = \psi_{n m}^{(\infty)} (x_\infty)\,
\prod_p \psi_{\alpha_p \beta_p}^{(p)} (x_p)\, , \quad x\in
{\mathcal A} \, , \eqno(2.17)
$$
where all but a finite number of $\psi_{\alpha_p \beta_p}^{(p)}
(x_p)$  are $\psi_{ 0 0}^{(p)} (x_p) = \Omega (|x_p|_p)\, $ [Eq.
(2.9)]. We have to take $\psi_{\alpha \beta} (x)$ in this form
because the Fourier transform of the vacuum state $\Omega
(|x_p|_p)$ is $\tilde{\Omega}  = \Omega (|k_p|_p)$. It enables
$\psi_{\alpha \beta} (x)$  and $\tilde{\psi}_{\alpha \beta} (k)$
to be functions of the adeles $x$ and $k$, where $k$ is a dual
(momentum) to $x$. Note that the Schwartz-Bruhat functions satisfy
this condition.

\bigskip
\noindent{\bf 3.\ \ Harmonic Oscillator over ${\mathbb R}$ and
${\mathbb Q}_p $}
\bigskip

The harmonic oscillator represents a very simple theoretical model
which can be solved exactly classically  as well as
quantum-mechanically.

\bigskip
\noindent{\bf  Case A:\ \ Ordinary and $p$-adic classical
oscillator}
\bigskip

The corresponding nonrelativistic classical Hamiltonian is
$$
H = \frac{1}{2 m} k^2 + \frac{m \omega^2}{2} q^2 \, , \quad m\neq
0 \, ,      \eqno(3.1)
$$
where $q$ and $k$ are the position and the momentum, respectively.
The classical time evolution of the phase space can be presented
in the form
$$
\Big( \begin{array} {c} q(t)  \\ k(t) \end{array}   \Big)  = T_t
\, \Big( \begin{array} {c} q  \\ k  \end{array} \Big)  = T_t \,
z\, , \, \, \, T_t = \Big( \begin{array} {lc} \cos{\omega t} &
(m\omega)^{-1}\, \sin{\omega t}  \\  - m \omega \sin{\omega t} &
\cos{\omega t}
\end{array}   \Big)\, ,  \eqno(3.2)
$$
where $q = q (0)$ and $k = k (0)$. In the real case all quantities
belong to ${\mathbb R}$. Analogously, in the $p$-adic classical
case they belong to ${\mathbb Q}_p$. Convergence domains for the
expansions of $p$-adic $\cos{\omega t}$ and $\sin{\omega t}$
require satisfying the conditions  $|\omega t|_p \leq p^{-1}$ for
$p\neq 2$ and $|\omega t|_2 \leq 2^{-2}$ for $p=2$. These domains
we denote by $G_p$, and they are additive groups. One can easily
show that $T_t\, T_{t'} = T_{t+t'}$ and $B (T_t z, T_t z') =
B(z,z')$, where
$$
B (z, z') =-k\, q' + q\, k'        \eqno(3.3)
$$
is the skew-symmetric bilinear form on the phase space.

\bigskip
\noindent{\bf  Case B: \ \ Ordinary quantum oscillator}
\bigskip

In ordinary quantum mechanics the harmonic oscillator is given by
the Schr\"odinger equation
$$
\frac{d^2 \psi^{(\infty)}}{d x^2} + \frac{2 m}{\hbar^2} \Big(
 E - \frac{m \omega^2}{2} x^2\Big) \psi^{(\infty)} = 0\, , \eqno(3.4)
$$
where $x, \, m, \, \omega, \, \hbar \in {\mathbb R}$ and
$\psi^{(\infty)} \in {\mathbb C}$. (For a function containing
index $\infty$ or $p$ we shall often omit such an index for its
variable.)

If we introduce a dimensional coordinate
$$
\xi = x \sqrt{2 \pi} \Big(\frac{m \omega}{h} \Big)^{\frac{1}{2}}
\, \eqno(3.5)
$$
Eq. $(3.4)$ becomes
$$
\frac{d^2 \psi^{(\infty)}}{d \xi^2} + \Big( \frac{4 \pi E}{h
\omega}
  - \xi^2\Big) \psi^{(\infty)} = 0\, . \eqno(3.6)
$$
from now on we make simplification using $m = \omega = h =1$.
Physical solutions to (3.6) represent  an orthonormal basis of an
$L_2 ({\mathbb R})$ and since $\xi = x\sqrt{2 \pi}$ we have
$$
\psi_n^{(\infty)} (x) =\frac{2^{\frac{1}{4}}}{(2^n
n!)^{\frac{1}{2}}}\, e^{-\pi x^2}\, H_n (x \sqrt{2 \pi}) \, ,
\quad n = 0,\, 1, \cdots ,    \eqno(3.7)
$$
where $H_n (x \sqrt{2 \pi})$ are the Hermite polynomials. One can
easily show that $\psi_n^{(\infty)} (x)$ satisfies the above
condition $(\it i)$ for an elementary function (2.8).

Ordinary quantum mechanics can also be given by a triple
\cite{volovich3,weyl}
$$
\big(L_2 ({\mathbb R}),\, W_\infty (z_\infty), \, U_\infty
(t_\infty) \big)\, ,      \eqno(3.8a)
$$
where $L_2 ({\mathbb R})$ is the Hilbert space defined in the
preceding section, $z_\infty$ is a point of real classical phase
space, $W_\infty (z_\infty)$ is a unitary representation of the
Heisenberg-Weyl group on  $L_2 ({\mathbb R})$, and $U_\infty
(t_\infty)$ is a unitary representation of the evolution operator
on $L_2 ({\mathbb R})$.

Recall that the Heisenberg-Weyl group consists of elements $(z,\,
\alpha)$ with the group product
$$
(z,\, \alpha)\, (z',\, \alpha) =\big( z + z',\, \alpha + \alpha' +
\frac{1}{2} B (z, z') \big) \, ,           \eqno(3.9)
$$
where $B (z, z')$ is as defined by (3.3) and $\alpha$ is a
parameter. The corresponding unitary representation in the real
case case is
$$
\chi_\infty (\alpha)\, W_\infty (z)\, ,          \eqno(3.10)
$$
and $W_\infty (z)$ satisfies the Weyl relation
$$
W_\infty (z)\, W_\infty (z') = \chi_\infty \Big(\frac{1}{2} B (z,
z')\Big)\, W_\infty (z+ z') \, .                 \eqno(3.11a)
$$
The operator  $W_\infty (z)$ acts on $\psi_n^{(\infty)} (x)$ in
the following way:
$$
W_\infty (z) \, \psi_n^{(\infty)} (x) = \chi_\infty \Big( -\frac{k
q}{2}  \Big)\, U_q \, \chi_\infty (k x) \, \psi_n^{(\infty)} (x)
$$
$$
= \chi_\infty \Big( \frac{k q}{2} + k x   \Big)\,
\psi_n^{(\infty)} (x + q)\, , \eqno(3.12a)
$$
where $U_q \psi (x) = \psi (x + q)$.

The evolution operator $U_\infty (t)$ is defined by
$$
U_\infty (t)\, \psi_n^{(\infty)} (x) =\int_{{\mathbb R}} {\mathcal
K}_t^{(\infty)} (x, y) \, \psi_n^{(\infty)} (y)\, dy \, ,
\eqno(3.13a)
$$
where the kernel $ {\mathcal K}_t^{(\infty)} (x, y)$ for the
harmonic oscillator is
$$
{\mathcal K}_t^{(\infty)} (x, y) = \lambda_\infty (2 \sin{t})\,
|\sin{t}|_\infty^{-\frac{1}{2}} \, \exp{2 \pi i \Big( \frac{x^2 +
y^2}{2 \tan{t}} -\frac{x y}{\sin{t}} \Big)}\, ,  \eqno(3.14a)
$$
$$
{\mathcal K}_0^{(\infty)} (x, y) = \delta_\infty (x -y) .
\eqno(3.13b)
$$
In (3.14a) the function $\lambda_\infty (a)$ is
$$
\lambda_\infty (a) = \frac{1}{\sqrt{2}}\, (1 - i\, \mbox{sign}\,
a) \, . \eqno(3.15a)
$$

The unitary operator $U_\infty (t)$ satisfies the group relation
$$
U_\infty (t+t')  = U_\infty (t)\, U_\infty (t')   \eqno(3.16a)
$$
and consequently for the kernel one has
$$
{\mathcal K}_{t+t'}^{(\infty)} (x, y) = \int_{{\mathbb R}}
{\mathcal K}_t^{(\infty)} (x, z)\,  {\mathcal K}_{t'}^{(\infty)}
(z, y) \, dz \, .            \eqno(3.17a)
$$
The operators $U_\infty (t)$ and $W_\infty (z)$ are connected by
the relation
$$
U_\infty (t)\, W_\infty (z)  = W_\infty (T_t z)\, U_\infty (t)\, .
\eqno(3.18a)
$$

Note that (3.7) are eigenfunctions of the evolution operator
$U_\infty (t)$, i.e.
$$
U_\infty (t) \psi_n^{(\infty)} (x) = \exp{(- 2 \pi i
E_n^{(\infty)}\, t)} \, \psi_n^{(\infty)} (x) \, , \eqno(3.19)
$$
where $E_n^{(\infty)} = \big( n + \frac{1}{2} \big)\, \frac{1}{2
\pi}$ is the corresponding energy. The right hand side of (3.19)
represents eigenfunctions of the energy operator $\hat{E} =
\frac{i}{2 \pi}\, \frac{\partial}{\partial t}$.

Let ${\mathcal D}_\infty$ be an observable and  ${\hat{\mathcal
D}}_\infty$ the corresponding operator which acts in $L_2 (
{\mathbb R})$. An expectation (average) value of ${\mathcal
D}_\infty$ in a state $\psi^{(\infty)} (x)$ is
$$
\langle {\mathcal D}_\infty \rangle = (\psi^{(\infty)}, \,
{\hat{\mathcal D}}_\infty \, \psi^{(\infty)}) \, .  \eqno(3.20a)
$$
It is also of interest to have a knowledge of the mean square
deviation $\Delta {\mathcal D}_\infty$, which is a measure of the
dispersion around $\langle{\mathcal D}_\infty\rangle$:
$$
\Delta {\mathcal D}_\infty = [ \langle({\mathcal D}_\infty
-\langle {\mathcal D}_\infty\rangle )^2\rangle]^{1/2} = ( \langle
{\mathcal D}_\infty^2\rangle - \langle {\mathcal
D}_\infty\rangle^2)^{1/2} \, . \eqno(3.21a)
$$
In the vacuum state
$$
\psi_0^{(\infty)} (x) = 2^{\frac{1}{4}} \, e^{-\pi x^2}\, , \quad
\tilde{\psi}_0^{(\infty)} (k) = 2^{\frac{1}{4}} \, e^{-\pi k^2}\,
,    \eqno(3.22)
$$
where  $\tilde{\psi}_0^{(\infty)} = {\psi}_0^{(\infty)}$ is the
Fourier transform of ${\psi}_0^{(\infty)}$, one obtains
$$
\langle x_\infty\rangle  =\langle k_\infty\rangle = 0\, ,
\eqno(3.23a)
$$
$$
\langle|x_\infty|_\infty^s \rangle = \langle|k_\infty|_\infty^s
\rangle = \sqrt{2} \Gamma \Big(\frac{s+1}{2}  \Big) \, (2
\pi)^{-\frac{s+1}{2}} \, ,   \eqno(2.23b)
$$
$$
\Delta x_\infty =  \Delta k_\infty = \frac{1}{2\sqrt{\pi}} \, ,
\quad \Delta x_\infty \Delta k_\infty = \frac{1}{4 \pi} \, ,
\eqno(3.23c)
$$
$$
\Delta |x_\infty|_\infty  = \Delta |k_\infty|_\infty =
\frac{1}{2\sqrt{\pi}} \Big( 1-\frac{2}{\pi} \Big)^{1/2}\, .
\eqno(3.23d)
$$

\bigskip
\noindent{\bf Case C: \ \ $p$-Adic quantum oscillator}
\bigskip

In $p$-adic quantum mechanics, which is the subject of this
subsection, canonical variables are $p$-adic numbers and wave
functions are complex-valued. Since wave functions and their
variables belong to differently valued number fields, the usual
(canonical) quantization does not work. However, one can make a
$p$-adic generalization of the above Weyl representation.

According to the Vladimirov-Volovich approach
\cite{volovich2,volovich3,volovich4}, $p$-adic quantum mechanics
is given by a triple
$$
\big( L_2 ({\mathbb Q}_p) \, , W_p (z_p)\, , U_p (t_p)   \big)\, ,
\eqno(3.8b)
$$
where $L_2 ({\mathbb Q}_p)$ is the $p$-adic Hilbert space defined
in the preceding section, $z_p$ is a point of $p$-adic classical
phase space, $W_p (z_p)$ is a unitary representation of the
Heisenberg-Weyl group on  $L_2 ({\mathbb Q}_p)$, and $U_p (t_p)$
is a $p$-adic  evolution operator which realizes  a unitary
representation on $L_2 ({\mathbb Q}_p)$ of a group $G_p$.

Analogously to the real case, the operator $W_p (z_p)$  is the
unitary representation of the Heisenberg-Weyl group  (3.9) with
$p$-adic values of $z$  and $\alpha$.
 It also satisfies the relations
$$
W_p (z)\, W_p (z') = \chi_p \Big(\frac{1}{2} B (z, z')\Big)\, W_p
(z+ z') \, ,                 \eqno(3.11b)
$$
$$
 W_p (z) \, \psi^{(p)}
(x) = \chi_p \Big( \frac{k q}{2} + k x   \Big)\, \psi^{(p)} (x +
q)\, . \eqno(3.12b)
$$
A $p$-adic evolution operator is given by
$$
U_p (t)\, \psi^{(p)} (x) =\int_{{\mathbb Q}_p} {\mathcal
K}_t^{(p)} (x, y) \, \psi^{(p)} (y)\, dy \, , \eqno(3.13b)
$$
where the kernel  for the harmonic oscillator is
$$
{\mathcal K}_t^{(p)} (x, y) = \lambda_p (2 t)\,
|t|_p^{-\frac{1}{2}} \, \chi_p{ \Big( \frac{x y}{\sin{t}}
-\frac{x^2 + y^2}{2 \tan{t}} \Big)}\, , \quad t\in G_p \setminus
\{ 0 \} \, , \eqno(3.24a)
$$
$$
{\mathcal K}_0^{(p)} (x, y) = \delta_p (x -y) \, , \eqno(3.24b)
$$
where $\delta_p (x - y)$ is a $p$-adic analog of the Dirac
$\delta$-function. For canonical expansion
$$
a = p^\nu \, (a_0 + a_1 p + a_2 p^2 + \cdots ) \, , \quad \nu \in
{\mathbb Z}\, , \quad a_0 \neq0\,\, , 0\leq a_i \leq p-1 \, ,
\eqno(3.25)
$$
the number-theoretic function $\lambda_p (a)$ is
$$
\lambda_p (a) = \left\{ \begin{array} {lll} 1\,,\quad & \nu = 2
k\,,
\quad & p\neq 2\,, \\
\big( \frac{a_0}{p} \big)\, , \quad & \nu = 2 k +1\, , \quad &
p\equiv
1(\mbox{mod}\,\, 4)\, ,  \\
i \big( \frac{a_0}{p} \big) \, , \quad & \nu = 2 k +1\, , \quad &
p\equiv 3(\mbox{mod}\,\, 4)\, ,
\end{array} \right.  \eqno(3.15b)
$$
$$
\lambda_2 (a) = \left\{ \begin{array} {ll} \frac{1}{\sqrt{2}}\,
[1+ (-1)^{a_1}\, i] \, , \quad & \nu = 2 k \, ,  \\
\frac{1}{\sqrt{2}}\, (-1)^{a_1 + a_2} \,  [1+ (-1)^{a_1}\, i] \, ,
\quad    & \nu = 2 k +1 \, ,
\end{array} \right.    \eqno(3.15c)
$$
where $\big( \frac{a_0}{p} \big)$ is the Legendre symbol and $k
\in {\mathbb Z}$.

The  operator $U_p (t)$ and its kernel ${\mathcal K}_{t}^{(p)} (x,
y) $ satisfy the group relations
$$
U_p (t+t')  = U_p (t)\, U_p (t')\, ,   \eqno(3.16b)
$$
$$
{\mathcal K}_{t+t'}^{(p)} (x, y) = \int_{{\mathbb Q}_p} {\mathcal
K}_t^{(p)} (x, z)\,  {\mathcal K}_{t'}^{(p)} (z, y) \, dz \, .
\eqno(3.17b)
$$

To prove (3.17b) we use
$$
\int_{{\mathbb Q}_p} \chi_p (\alpha x^2 + \beta x)\, dx =
\lambda_p (\alpha) \, |2 \, \alpha|_p^{-\frac{1}{2}} \, \chi_p
\Big( - \frac{\beta^2}{4 \alpha}\Big) \, , \quad \alpha \neq 0 \,
,
$$
with properties of $\lambda_p (a)$: $ \lambda_p (0) = 1, \,\,
\lambda_p (a^2 b) = \lambda_p (b) , \,\,  \lambda_p (a)\,
\lambda_p (b) =  \lambda_p (a+ b)\, \lambda_p (a^{-1} + b^{-1})$.
The operators $U_p (t)$ and $W_p (z)$ are connected in the same
way as in the real case:
$$
U_p (t)\, W_p (z)  = W_p (T_t z)\, U_p (t)\, . \eqno(3.18b)
$$

A spectral problem in $p$-adic quantum mechanics is related to an
investigation of the eigenvalues and eigenfunctions of the
evolution operator. According to Ref. 11, a character $\chi_p
(\alpha  t)$ is an eigenvalue  of $U_p (t)$ for the harmonic
oscillator if and only if $\alpha$ takes one of the values
$$
\alpha = 0 \, ,               \eqno(3.26a)
$$
$$
\alpha = p^{-\nu} (\alpha_0 + \alpha_1 p +\cdots + \alpha_{\nu -
2} p^{\nu-2}) \, ,\quad \alpha_0 \neq 0\, , \quad 0\leq \alpha_i
\leq p-1\, ,    \eqno(3.26b)
$$
where $(\it i)$ $\nu \geq 2$ for $p \equiv 1 (\mbox{mod}\,\, 4) \,
,$ $\,\, (\it ii) \,\,  \nu = 2 n \,\, (\nu \in {\mathbb N})$ for
$p \equiv 3 (\mbox{mod}\,\, 4) \ ,$ and $(\it iii) \,\, \nu \geq 3
\,\,$  for $p = 2\,\, $ [with $ \alpha_{\nu -3} p^{\nu -3}$  as a
last term in (3.26b)]. The set of $\alpha$ in Eqs. (3.26) we
denote by $I_\alpha$. We can introduce
$$
E_{\alpha}^{(p)} = \alpha_p \, , \quad \alpha_p \in I_p \, ,
\eqno(3.27)
$$
which may be regarded as a discrete $p$-adic energy of the
harmonic oscillator.

The corresponding eigenstates satisfy the equation
$$
U_p (t)\, \psi_{\alpha \beta}^{(p)} (x) = \chi_p (\alpha\, t) \,
\psi_{\alpha \beta}^{(p)} (x) \, , \eqno(3.28)
$$
where the index $\beta$ differentiates eigenfunctions for
degenerate states. In particular, $\alpha = 0$ corresponds to a
vacuum state which is invariant under $U_p (t)$, i.e.
$$
U_p (t)\, \psi_{0 \beta}^{(p)} (x) =   \psi_{0 \beta}^{(p)} (x) \,
. \eqno(3.29)
$$

The Hilbert space $L_2 ({\mathbb Q}_p)$ can be presented as
$$
L_2 ({\mathbb Q}_p) =\bigoplus_{\alpha \in I_p} \, H_\alpha^{(p)}
\, , \eqno(3.30)
$$
which is a direct sum of mutually orthogonal subspaces
$H_\alpha^{(p)}$. The dimensions of $H_\alpha^{(p)}$ are as
follows:

$(\it i)$ When $p\equiv 1 (\mbox{mod}\, \, 4)$, dim
$H_\alpha^{(p)} = \infty $ for every  $\alpha \in I_p$;

$(\it ii)$ When $p \equiv 3 (\mbox{mod}\,\, 4)$, dim $H_0^{(p)}
=1$ but dim $H_\alpha^{(p)} = p+1$ for $|\alpha|_p \geq p^{2 n} \,
\, (n \in {\mathbb N})$;

$(\it iii)$ When $p = 2$, dim $H_0^{(2)}  = $ dim $H_\alpha^{(2)}
= 2$ for $|\alpha|_2 = 2^3$ and dim $H_\alpha^{(2)} = 4$  for
$|\alpha|_2 \geq 2^4 $. These dimensions determine the number of
linearly independent eigenfunctions for any eigenvalue $\chi_p
(\alpha t)$. The eigenfunctions $\psi_{\alpha \beta}^{(p)} (x)$
are obtained \cite{volovich5,zelenov2} in an explicit form for the
vacuum state $(\alpha =0)$ and for some excited states $\alpha
\neq 0$. All $\psi_{\alpha \beta}^{(p)} (x)$ satisfy the condition
$(\it ii)$ of the elementary functions. The orthonormal vacuum
eigenfunctions of $U_p (t)$ for the harmonic oscillator are:

$(\it i) \mbox{For}\,\, p\equiv 1 (\mbox{mod} \,  \, 4)\, ,\,
\psi_{0 0}^{(p)} (x) = \Omega (|x|_p)\, , \, \, \psi_{0 \nu}^{(p)}
(x) $

$ =p^{-\frac{\nu}{2}}\, (1 - p^{-1})^{-\frac{1}{2}}\, \chi_p{(\tau
x^2)}  \, \delta (p^\nu - |x|_p)\, , \, \, \nu \in {\mathbb N}, \,
\tau^2 = -1 \, ;  \quad  \quad \quad (3.31a)$

$(\it ii) \mbox{For} \,\, p\equiv 3 (\mbox{mod}\, \, 4)\, , \,
\psi_{0 0}^{(p)} (x) =\Omega (|x|_p) \,; \qquad \qquad \qquad
\qquad \qquad  (3.31b)$

$(\it iii) \mbox{For}\,\, p = 2 \, ,\, \psi_{0 0}^{(2)} (x)
=\Omega (|x|_2) \, , \, \psi_{0 1}^{(2)} (x) =2 \, \Omega (2
|x|_2) - \Omega (|x|_2) \, . \,    (3.31c)$

In (3.31a) $\delta (p^\nu -|x|_p)$ is an elementary function
defined  by
$$
\delta (p^\nu -|x|_p) =\left\{\begin{array}{ll} 1\, , \quad &
|x|_p = p^\nu \, , \\
0\,, \quad & |x|_p \neq p^\nu .
\end{array} \right.    \eqno(3.32)
$$

One can generalize (3.20a) and (3.21a) to the $p$-adic case. Thus
one obtains
$$
\langle {\mathcal D}_p \rangle = (\psi^{(p)}, \, {\hat{\mathcal
D}}_p \, \psi^{(p)}) \, ,   \eqno(3.20b)
$$
$$
\Delta {\mathcal D}_p = [ \langle({\mathcal D}_p -\langle
{\mathcal D}_p\rangle )^2\rangle]^{1/2} = ( \langle {\mathcal
D}_p^2\rangle - \langle {\mathcal D}_p\rangle^2)^{1/2} \, .
\eqno(3.21b)
$$
Note that $\langle x_p \rangle$  and $\langle k_p \rangle$ have no
meaning. For the simplest vacuum state
$$
\psi_{0 0}^{(p)} (x) = \Omega (|x|_p)\, , \tilde{\psi}_{0 0}^{(p)}
(x) = \Omega (|k|_p) \, ,         \eqno(3.33)
$$
where $ \tilde{\psi}_{0 0}^{(p)} = \psi_{0 0}^{(p)} $ is the
Fourier transform of $\psi_{0 0}^{(p)} (x)$, we get
$$
\langle|x|_p^s\rangle  =\langle|k|_p^s\rangle = \frac{1-p^{-1}}{1
- p^{-s-1}}\, , \quad \mbox{Re}\, s > -1 \, ,    \eqno(3.34a)
$$
$$
\Delta |x|_p =\Delta |k|_p = \Big( \frac{1-p^{-1}}{1-p^{-3}}
\Big)^{\frac{1}{2}} \, \Big[ 1 -
\frac{(1-p^{-1})(1-p^{-3})}{(1-p^{-2})^2} \Big]^{\frac{1}{2}} \, .
\eqno(3.34b)
$$

\bigskip
\noindent{\bf 4.\ \  Harmonic Oscillator over Adeles }
\bigskip

Let us define adelic quantum mechanics as a triple
$$
\big( L_2 ({\mathcal A}) \, , W (z)\, , U (t)   \big)\, ,
\eqno(4.1)
$$
where  ${\mathcal A}$ is the additive group of adeles, $z$ is an
adelic point of a classical phase space, and $t$ is an adelic
time. $L_2 ({\mathbb Q}_p)$ is the  Hilbert space of
complex-valued square integrable functions on ${\mathcal A}$ (see
Sec. 2), $W (z)$ is a unitary representation of the
Heisenberg-Weyl group on $L_2 ({\mathcal A})$, and $U (t)$ is a
unitary representation of the evolution operator on $L_2
({\mathcal A})$.

The Heisenberg-Weyl group (3.9) is generalized to the adelic case
by taking $z \in {\mathcal A}^2 = {\mathcal A} \times {\mathcal
A}, \quad \alpha \in {\mathcal A}, $ and $B (z,z') \in {\mathcal
A}$.
 In this case the unitary representation of (3.9) is
 $$
\chi (\alpha)\, W (z) = \chi_\infty (\alpha_\infty) \, W_\infty
(z_\infty) \, \prod_p \chi_p (\alpha_p) \, W_p (z_p) \, ,
\eqno(4.2)
 $$
where $\chi$ is defined by (2.6), and the Weyl relation
$$
W (z) \, W(z') = \chi \Big(\frac{1}{2}B(z,z')\Big)\, W (z+z')
\eqno(4.3)
$$
is satisfied. On the basis of (3.12a) and (3.12b) we have
$$
W (z) \, \psi (x) = \chi \Big(\frac{k q}{2} +  kx \Big) \, \psi
(x+q). \eqno(4.4)
$$
When $x, \, q,\, k$ are principal adeles (adelic rational
variables) then $\chi \big(\frac{k q}{2} +  k x \big) = 1$ and
$$
W (z) \, \psi (x) = \psi (x +q) .    \eqno(4.5)
$$

Let the evolution operator $U (t)$ be defined by
$$
U (t)\, \psi (x) = \int_{\mathcal A} {\mathcal K}_t (x, y) \, \psi
(y) \, dy \, ,        \eqno(4.6)
$$
where $t \in G \subset {\mathcal A}, \quad x, y \in {\mathcal A},$
and $ \psi (x) \in L_2 ({\mathcal A})$. Also $\, U (t) = U_\infty
(t_\infty)\, \prod_p U_p (t_p)$ and
$$
{\mathcal K}_t (x, y) = {\mathcal K}_{t_\infty}^{(\infty)}
(x_\infty, y_\infty)\, \prod_p {\mathcal K}_{t_p}^{(p)} (x_p,
y_p)\, , \eqno(4.7)
$$
where ${\mathcal K}_{t_\infty}^{(\infty)} (x_\infty, y_\infty)$
and ${\mathcal K}_{t_p}^{(p)} (x_p , y_p)$ for the harmonic
oscillator are given by (3.14) and (3.24). Denoting $\lambda (a) =
\lambda_\infty (a_\infty) \, \prod_p \lambda_p (a_p) $ one can
write
$$
{\mathcal K}_t (x, y) = \lambda (2 \sin{ t})\,
|\sin{t}|^{-\frac{1}{2}} \, \chi{ \Big( \frac{x y}{\sin{t}}
-\frac{x^2 + y^2}{2 \tan{t}} \Big)}\, ,   \eqno(4.8)
$$
which resembles the form of real and $p$-adic kernels. The
infinite products (4.7) and (4.8) are divergent, and they
represent generalized functions which make definite sense under
adelic integration. By virtue of (3.16a), (3.16b) and (3.17a),
(3.17b), $U (t)$ and ${\mathcal K}_t (x, y)$ satisfy group
relations
$$
U (t + t') = U (t)\, U(t') ,      \eqno(4.9)
$$
$$
{\mathcal K}_{t+t'} (x, y) = \int_{\mathcal A} {\mathcal K}_{t}
(x, z)\, {\mathcal K}_{t'} (z, y) \, dz \, .   \eqno(4.10)
$$

Note that (4.10) does not represent a product of (3.17a), (3.17b)
with integration over the whole ${\mathbb Q}_p$ for every $p$.
Such product would be inconsistent with the adelic approach.
Adelic integration means that (4.10) consists of
$$
{\mathcal K}_{t+t'}^{(p)} (x, y) = \int_{|z|_p \leq 1 } {\mathcal
K}_{t}^{(p)} (x, z)\, {\mathcal K}_{t'}^{(p)} (z, y) \, dz \,
\quad  x, y \in {\mathbb Z}_p \, , \eqno(4.11)
$$
for all but a finite number of $p$. This restricted integration is
in a close connection with the vacuum state $\Omega (|x|_p)$.
Equation (4.11) can be derived using the Gauss integral
\cite{volovich5}
$$
\int_{|x|_p \leq 1} \chi_p (\alpha x^2 + \beta x) \, dx =
\lambda_p (\alpha) \, |2 \alpha|_p^{-\frac{1}{2}} \, \chi_p \Big(
-\frac{\beta^2}{4 \alpha} \Big) \, \Omega \Big( \Big|
\frac{\beta}{2 \alpha} \Big|_p \Big) \, , \, \, \, |\alpha|_p > 1
\, . \eqno(4.12)
$$
The group $G$ for the adelic harmonic oscillator is
$$
G = {\mathbb R} \times G_2 \times \cdots \times G_p \times \cdots
\, ,
$$
where $G_2 = \{ t \in {\mathbb Q}_2 \, : |t|_2 \leq 2^{-2}  \}$
and $G_p = \{ t \in {\mathbb Q}_p \, : |t|_p \leq p^{-1}  \}$ for
$p\neq 2$. One can easily see that an adelic time cannot be a
principal idele.

Using (3.18a), (3.18b) and factorization of $U (t)$ and $W (z)$ on
real and $p$-adic parts, one can show that
$$
U (t) \, W (z) = W (T_t z)\, U (t) \, .          \eqno(4.13)
$$

By virtue of (3.19) and (3.28) the eigenfunctions $\psi_{\alpha
\beta}$ of the adelic harmonic oscillator must satisfy the
equation
$$
U (t) \, \psi_{\alpha \beta} (x) = \chi (E t)\, \psi_{\alpha
\beta} (x)\, ,                          \eqno(4.14)
$$
where $\alpha$ and $\beta$ are adelic indices of the form
$$
\alpha  =(n, \, \alpha_2 , \cdots , \alpha_p , \cdots) \, ,
\eqno(4.15a)
$$
$$
\beta = (0, \, \beta_2 , \cdots , \beta_p , \cdots) \, .
\eqno(4.15b)
$$

According to the definition (2.17), it follows that any eigenstate
of the adelic harmonic oscillator is
$$
\psi_{\alpha \beta} (x) = \frac{2^{\frac{1}{4}}}{(2^n n!
)^{\frac{1}{2}}} \, e^{- \pi x^2} \, H_n (x\sqrt{2 \pi}) \,
\prod_{p \in \Gamma_{\alpha \beta}} \psi_{\alpha_p \beta_p} (x_p)
\, \prod_{p \notin \Gamma_{\alpha \beta}} \Omega (|x_p|_p)\, ,
\eqno(4.16)
$$
where $\Gamma_{\alpha \beta}$ is a finite set of the primes for
which at least one of the indices $\alpha_p$ and $\beta_p$ is
different from zero. It is obvious that any eigenstate  (4.16) is
a Schwartz-Bruhat function. Any finite linear combination  of
these eigenstates is also a Schwartz-Bruhat function. The
eigenfunctions (4.16) are orthonormal and any $\psi (x) \in L_2
({\mathcal A})$ can be presented as
$$
\psi (x)  = \sum C_{\alpha \beta} \, \psi_{\alpha \beta} (x)\, ,
\eqno(4.17)
$$
where $C_{\alpha \beta} = (\psi_{\alpha \beta} , \psi)$ and $\sum
|C_{\alpha , \beta}|_\infty^2 = 1.$

According to (3.20a), (3.20b) and (3.21a), (3.21b) we define
$$
\langle {\mathcal D} \rangle = (\psi , \, {\hat{\mathcal D}} \,
\psi ) \, , \quad \psi \in L_2 ({\mathcal A })\, ,   \eqno(4.18)
$$
$$
\Delta {\mathcal D} = [ \langle({\mathcal D} -\langle {\mathcal D}
\rangle )^2\rangle]^{1/2} = ( \langle {\mathcal D}^2\rangle -
\langle {\mathcal D} \rangle^2)^{1/2} \, , \eqno(4.19)
$$
where $\hat{\mathcal D} = \hat{\mathcal D}_\infty \,  \prod_p
\hat{\mathcal D}_p$ is the adelic operator of the observable
${\mathcal D}$. For a normalized state (4.17) we have
$$
\langle {\mathcal D} \rangle = \sum |C_{\alpha \beta}|_\infty^2 \,
\langle {\mathcal D} \rangle_{\alpha \beta} \, ,
$$
$$
\langle {\mathcal D} \rangle_{\alpha \beta} =  (\psi_{\alpha
\beta} , \, {\hat{\mathcal D}} \, \psi_{\alpha \beta} )    =
\langle {\mathcal D}_\infty \rangle_n \prod_p \langle {\mathcal
D}_p \rangle_{\alpha_p \beta_p} \, .      \eqno(4.20)
$$

Let us denote
$$
\psi_{0 0} (x) = \psi_0^{(\infty)} (x_\infty) \, \prod_p \psi_{0
0}^{(p)} (x_p) = 2^{\frac{1}{4}}\, e^{-\pi x_\infty^2} \prod_p
\Omega (|x_p|_p)\, ,          \eqno(4.21a)
$$
$$
|x|_{(p_n)}^s = |x_\infty|_\infty^s \prod_{p=2}^{p_n} |x_p|_{p}^s
\, , \quad   |k|_{(p_n)}^s = |k_\infty|_\infty^s \prod_{p=2}^{p_n}
|k_p|_{p}^s \, ,                       \eqno(4.22)
$$
where $s \in {\mathbb C}$ and $p_n$ is an $n$th prime. As a
consequence of (3.22) and (3.33) the Fourier transform of (4.21a)
is
$$
\tilde{\psi}_{0 0} (k) =  2^{\frac{1}{4}}\, e^{-\pi k_\infty^2}
\prod_p \Omega (|k_p|_p)  \, .       \eqno(4.21b)
$$
Expectation values of (4.22) in the simplest vacuum state $\psi_{0
0}$ are
$$
\langle |x|_{(p_n)}^s \rangle = \langle |k|_{(p_n)}^s \rangle =
\sqrt{2} \,\, \Gamma\Big(\frac{s+1}{2} \Big) \, (2 \pi)^{-
\frac{s+1}{2}} \, \prod_{p=2}^{p_n} \frac{1- p^{-1}}{1 - p^{-s-1}}
 \, . \eqno(4.23)
$$
For the corresponding $(s = 1)\, \, \, \Delta |x|_{(p_n)}$ and
$\Delta |k|_{(p_n)}$ one gets
$$
\Delta |x|_{(p_n)} = \Delta |k|_{(p_n)} =  \Big( \frac{1}{4 \pi}\,
\prod_{p=2}^{p_n} \frac{1 -p^{-1}}{1 - p^{-3}} \Big)^{\frac{1}{2}}
\, \left[1 - \frac{2}{\pi} \, \prod_{p=2}^{p_n}
\frac{(1-p^{-1})(1-p^{-3})}{(1-p^{-2})^2} \right]^{\frac{1}{2}}  .
\eqno(4.24)
$$
When $p_n \rightarrow \infty$ one obtains
$$
\langle |x|^s\rangle =\langle |k|^s\rangle =  \Delta |x| = \Delta
|k|  = 0 \, .     \eqno(4.25)
$$

Note that the vacuum state $\psi_{0 0 } (x)$ [Eq. (4.21a)] is the
simplest elementary function (2.8) defined on adeles. Applying the
Mellin transform (2.11) to $\psi_{0 0} (x)$ one gets
$$
\Phi (s) = \sqrt{2}\,\, \Gamma \Big( \frac{s}{2}\Big) \,
\pi^{-\frac{s}{2}}  \, \zeta (s) \, .    \eqno(4.26)
$$
Since $\tilde{\psi}_{00} = \psi_{0 0}$ it follows that
$\tilde{\Phi} (s) = \Phi (s)$. If we replace $\Phi $ by
$\tilde{\Phi}$ in the Tate formula (2.12) by (4.26), there appears
the well-known functional relation for the Riemann zeta function:
$$
\pi^{-\frac{s}{2}}\, \Gamma \Big( \frac{s}{2}\Big)  \, \zeta (s) =
\pi^{\frac{s-1}{2}}\, \Gamma \Big( \frac{1- s}{2}\Big)  \, \zeta
(1- s)   \, .                               \eqno(4.27)
$$

The expectation value of $U (t)$ in an adelic eigenstate (4.16) is
equal to its eigenvalue:
$$
U (t) = \chi (E t) = e^{2 \pi i \{ E t \}}    \eqno(4.28)
$$
where
$$
\{   E t \} = - E_\infty\, t_\infty + \sum_p \{ E_p \,t_p \}_p
\eqno(4.29)
$$
is an adelic dynamical phase.

\bigskip
\noindent{\bf 5.\ \ Discussion and Concluding Remarks}
\bigskip

For one-dimensional dynamical systems, whose classical time
evolution leaves invariant the symplectic bilinear form (3.3), we
have formulated adelic quantum mechanics. As an illustration we
considered the corresponding model of a harmonic oscillator. Let
us discuss this model and some general features of quantum
mechanics on adeles.

The adelic harmonic oscillator is a natural illustration of
mathematical analysis over adeles. It is a simple, exact and
instructive adelic model. All eigenstates, as well as all their
finite linear combinations, are Schwartz-Bruhat functions. The
simplest vacuum state is also the simplest elementary function on
adeles and its form is invariant under the Fourier transformation.
The Mellin transform of this state is the same function on
configuration and momentum space.

Recall that ordinary and $p$-adic quantum mechanics are theories
on real and $p$-adic spaces, respectively.  Adelic quantum
mechanics may be regarded as a theory on adelic space ${\mathcal
A}$.

One may generalize the concept of real matter, and introduce
$p$-adic and adelic matter, as well. Real and $p$-adic particles
belong to different (real and $p$-adic) parts of the adelic space
and they do not mutually interact. They are distinct low energy
limits of the adelic matter which exists at very high energies.
Adelic particles are unstable and decay to real and $p$-adic ones,
which belong to their own spaces.

According to the above assumption adelic quantum theory for real
particles should be equivalent to ordinary quantum theory. This is
really the case for the harmonic oscillator. For example,
according to (4.21a), (4.21b) and $\Omega (|0|_p) = 1$, the
simplest adelic vacuum state for a real harmonic oscillator $x_p
=0$ becomes the corresponding vacuum state within ordinary quantum
mechanics.

One might get impression that in solving the adelic problem one
has to collect real and $p$-adic solutions and take their product.
It is not quite correct. Namely, when adelic solution is a product
it is a very restricted one. One has also to take care of
convergence and adelic consistence. Moreover, there are solutions
which are not a product of real and $p$-adic parts [see e.g.
(4.24)].

By virtue of the above discussion the adelic harmonic oscillator
is a system of two particles whose interaction in ${\mathcal A}$
is presented by the potential
$$
V (x) = \Big( \frac{x_\infty^2}{2} , \, \frac{x_2^2}{2} , \cdots ,
\frac{x_p^2}{2} , \cdots \Big) \, , \quad m=\omega = 1 \, .
$$

Uncertainty relations in the simplest vacuum state are given by
(3.23c), (3.23d) for the real coordinates, and by (3.34b) for the
$p$-adic ones. We also calculated some expectation values of the
product of norms for position and momentum.

Finally, the above nonrelativistic mode may be regarded as a first
step towards a more profound relativistic adelic quantum theory.

\end{document}